\documentclass[aps,pra,superscriptaddress,twocolumn]{revtex4-2}
\usepackage{bm}
\usepackage{amsmath}
\usepackage{amsfonts}
\usepackage{amssymb}
\usepackage{graphicx}
\usepackage{color}
\usepackage{xcolor}
\usepackage[colorlinks,linkcolor=blue,anchorcolor=blue,citecolor=blue,urlcolor=blue]{hyperref}
\usepackage{dcolumn}
\usepackage{fancyhdr}

\begin{document}

\title{Detection of roton and phonon excitations in a spin-orbit coupled Bose-Einstein condensate with a moving barrier}
	
\author{Hao~Lyu}
\affiliation{Quantum Systems Unit, Okinawa Institute of Science and Technology Graduate University, Onna, Okinawa 904-0495, Japan}

\author{Yongping~Zhang}
\affiliation{Department of Physics, Shanghai University, Shanghai 200444, China}
	
\author{Thomas~Busch}
\affiliation{Quantum Systems Unit, Okinawa Institute of Science and Technology Graduate University, Onna, Okinawa 904-0495, Japan}

\begin{abstract}
		
We propose to detect phonon and roton excitations in a two-dimensional Bose-Einstein condensate with Raman-induced spin-orbit coupling
by perturbing the atomic cloud with a weak barrier. 
The two excitation modes can be observed by moving the barrier along different directions in appropriate parameter regimes.
Phonon excitations are identified by the appearance of solitary waves, while roton excitations lead to distinctive spatial density modulations. 
We show that this method can also be used to determine the anisotropic critical velocities of superfluid.  

\end{abstract}
	
\maketitle

\section{Introduction}
	
Superfluids possess the property of allowing for dissipationless flow around a barrier as long as the velocity of the flow does not exceed a critical value. This value can famously be determined from the slope of the excitation spectrum~\cite{Landau1,Landau2}, which usually can be separated into two regions allowing for collective excitations of phonon and roton types. Phonons exhibit a linear dispersion near the zero momentum, and the associated critical velocity is the so-called phonon velocity. Rotons, on the other hand, have a parabolic dispersion at a finite momentum and their presence in the excitation spectrum is known to lead to critical velocities lower than the phonon velocity. This scenario has been experimentally confirmed in liquid helium by moving an obstacle through the stationary superfluid~\cite{Allum,Castelijns}.

A more tunable platform that allows one to investigate the physics of superfluids is the atomic Bose-Einstein condensate (BEC)~\cite{Dalfovo,Leggett}. For these systems 
the phonon critical velocity has been measured with the help of a moving barrier, which can be realized in this case by a blue detuned laser~\cite{Raman,Onofrio,Engels}. In fact, the moving-barrier method is widely used to detect various different excitation modes and measure critical velocities in different systems and situations~\cite{Neely,Wilson,Wouters,Ramanathan,Desbuquois,Weimer,Singh,Kim}. For standard BECs with point-like particle interactions this process excites phonon modes in the atomic cloud which manifest in the appearance of density modulations and solitary waves~\cite{Engels}. 
Roton excitations can emerge in these systems if the dispersion relation is engineered to possess a double-well shape and in one recent experiment this was achieved by loading a BEC into a shaken optical lattice and dragging a speckle pattern through it~\cite{Chin}. 
In this experiment it was also confirmed that the critical velocities in the roton and non-roton directions are different, as a result of the asymmetry of the excitation spectrum.
Recently, anisotropic superfluids were also demonstrated in dipolar BECs~\cite{Ticknor,Wenzel}, which also possess roton excitations.

\begin{figure*}[tb]
\includegraphics[width=6in]{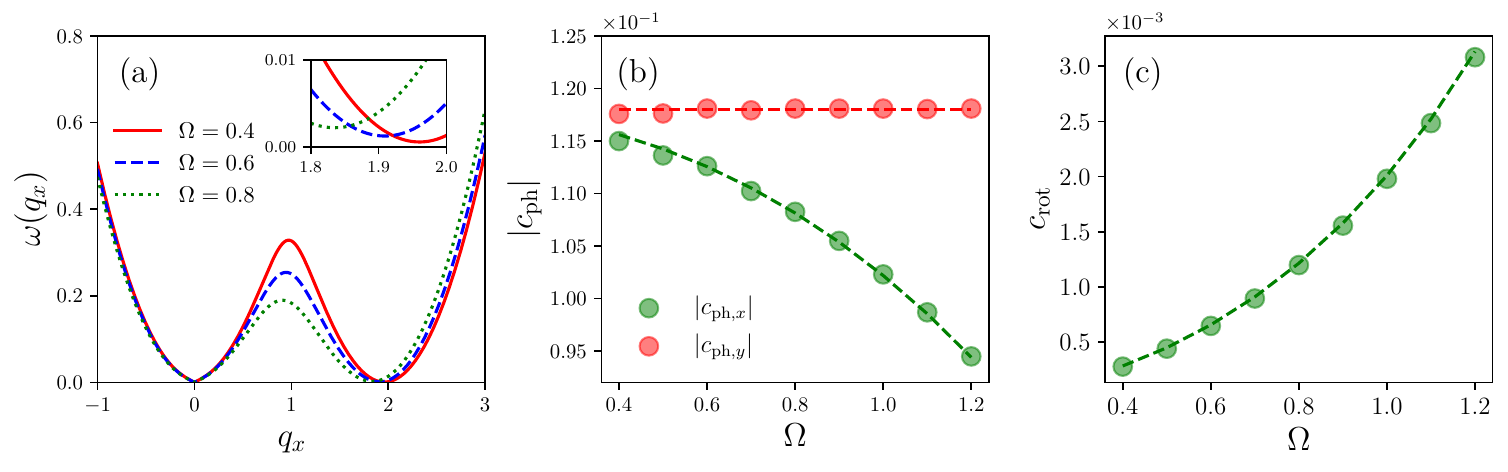}
\caption{Collective excitation spectrum and critical velocities of a two-dimensional BEC with Raman-induced spin-orbit coupling along the $x$-direction. 
(a) Excitation spectrum for various $\Omega$. The inset in (a) shows the roton gap.
(b) Critical velocity of the phonon modes along the $x$ (green dots) and $y$ (red dots) directions. 
(c) Critical velocity of the roton mode. In both panels the dots represent critical velocities calculated from the excitation spectrum of the homogeneous system, while the dashed lines are calculated directly from Eqs.~(\ref{eq:phonon}) and (\ref{eq:roton}). 
The atom density is set as $n=0.025$.}
\label{vc-BdG}
\end{figure*}
	
Another route to realize a double-well shaped dispersion that allows for roton excitations is synthetic spin-orbit coupling~\cite{Goldman,Zhai2015,Zhang}.
Spin-orbit coupling can be induced in ultracold atoms by using a pair of Raman lasers to couple two hyperfine states of the atoms which act as pseudospin states~\cite{Lin,Cheuk,Wang}. 
Spin-orbit coupled BECs have been studied intensively in recent years as, since the Galilean covariance is broken in such a system,
they possess various interesting physical  phenomena~\cite{ZhuQ,HeP,Wu,Campbell,Khamehchi,LiJR,Kato,Hou,Yi,ChenX}. 
Phonon and roton structures have been measured by using the Bragg spectroscopy~\cite{Ji,Khamehchi2014} and 
collective excitations have also been observed by sweeping a moving barrier through the spin-orbit coupled BEC \cite{Khamehchi2014}.
However, the effect of the different excitation modes on the density distribution is still not fully explored.
In addition, the superfluid critical velocities and their dependence on the propagation directions of the quasiparticles, have not been measured directly by the moving-barrier method in such a system.

In this work, we explore how to detect phonon and roton excitations in a two-dimensional BEC with Raman-induced spin-orbit coupling by using a moving barrier.
We show that different density-modulated states can be observed when phonon and or roton modes are excited by a barrier with appropriately chosen parameters, which is different to what can be done with Bragg spectroscopy.
The critical velocities can also be measured by this method and
our results provide an alternative way to detect excitations and explore dynamics in spin-orbit coupled condensates~\cite{Mossman,ZhuQL,LiYan}.
The structure of the manuscript is as follows. 
In Sec.~\ref{Sec:II} we present the critical velocities obtained from the excitation spectrum of a homogeneous system and in Sec.~\ref{Sec:III} the different excitation modes and critical velocities are studied by using a moving barrier.
The mismatch between the critical velocities calculated from the excitation spectra and moving-barrier method is discussed and the conclusion are presented in Sec.~\ref{Sec:IV}.

\section{Critical velocities of the homogeneous system}
\label{Sec:II}
	
We consider a $^{87}$Rb BEC with spin-orbit coupling along the $x$ direction, 
which can be realized by coupling two hyperfine states of the atoms with a pair of Raman beams~\cite{Lin,Ji}.
For $^{87}$Rb atoms, the differences between the spin-dependent interactions are very small~\cite{Lin} and one can approximately consider the different interactions strengths to be spin-independent.
The interaction coefficients can therefore be written as $g_{ij}=g\equiv 4\pi\hbar^2 a_s/m$ ($i,j=1,2$), where 
the $s$-wave scattering length is $a_s\approx100a_0$ with $a_0$ being the Bohr radius, and $m$ is the atom mass. 
For simplicity, we initially assume that the system is homogeneous.
	
The mean-field energy functional of the spin-orbit coupled spin-1/2 BEC is given by
\begin{align}
\mathcal{E}[\psi]&=\int d\bm{r}\psi^\dagger(\bm{r})H_{\mathrm{SOC}} \psi(\bm{r})  \notag\\
&\phantom{={}}+\frac{g}{2}\sum_{i,j=1,2}\int d\bm{r}|\psi_i(\bm{r})|^2|\psi_j(\bm{r})|^2,
\label{eq:energy}
\end{align}
where $\psi=(\psi_1,\psi_2)$ is the spinor wave function. 
The spin-orbit coupled Hamiltonian for a single particle can be written as
\begin{align}
H_{\mathrm{SOC}}=-\frac{1}{2}\sigma_0\nabla^2-i\frac{\partial}{\partial x}\sigma_z+\frac{\Omega}{2}\sigma_x,
\end{align}
with $\Omega$ being the Rabi frequency of the Raman lasers which depends on the laser intensity. 
$\sigma_{x,z}$ are the standard Pauli matrices and $\sigma_0$ is the two-dimensional identity matrix.
In our dimensionless calculations the units of length, momentum, frequency, and energy are chosen as $1/k_{\mathrm{Ram}}$, $\hbar k_{\mathrm{Ram}}$, $\hbar k^2_{\mathrm{Ram}}/m$, and $\hbar^2 k^2_{\mathrm{Ram}}/m$, respectively, where $ k_{\mathrm{Ram}}=2\pi/\lambda_{\mathrm{Ram}}$ with $\lambda_{\mathrm{Ram}}$ being the wavelength of the Raman lasers.
	
Spin-orbit coupled BECs exhibit a rich phase diagram that can be accessed by tuning the relative parameters~\cite{Lin,Martone,LiY}. 
In this work we only consider the plane-wave phase in which the $Z_2$ symmetry is broken and which, for appropriately chosen parameters~\cite{Zhang}, has a double-well shaped single-particle dispersion. The system therefore supports the excitation of roton modes, however only in one specific direction, while phonon modes can be excited in any directions~\cite{Yu}.
The different energy-momentum relations of phonon and roton modes lead to anisotropic critical velocities, which have already been observed in experiments~\cite{Khamehchi2014}.

A suitable ansatz for the ground-state wave functions of the plane-wave phase is given by $\psi_i(\bm{r})=\sqrt{n}\varphi_i e^{ikx}$ with the normalization condition $|\varphi_1|^2+|\varphi_2|^2=1$. Here $k$ is the quasimomentum of the plane wave, and $n$ is the density of the atoms. The parameters $\varphi_{1,2}$ and $k$ can be determined by substituting the ansatz into Eq.~(\ref{eq:energy}) and then minimizing the resulting free energy functional $F=\mathcal{E}-\mu N$, where $\mu$ is the chemical potential and $N$ is the number of atoms. This leads to the stationary Gross-Pitaevskii (GP) equations
\begin{align}	
\mu\psi= \left( H_{\mathrm{SOC}}+ng\sigma_0\right) \psi,
\label{eq:GPE}
\end{align}
and the collective excitation spectrum can be calculated once the ground state and chemical potential are known. 
Here, the units of the density $n$ and interaction strength $g$ are $k^3_{\mathrm{Ram}}$ and $\hbar^2/mk_{\mathrm{Ram}}$, respectively.
By considering the system to be perturbed by a weak external field, one can write the total wave function as a sum of the ground-state wave function and perturbation terms as
\begin{align}
\Psi_{1,2}=e^{-i\mu t}\left[\psi_{1,2}+u_{1,2}(\bm{r})e^{-i\omega t}+v_{1,2}(\bm{r})e^{i\omega t}\right],
\end{align}
where $\omega$ is the excitation energy. The amplitudes of the perturbations are given by $u_{i}(\bm{r})$ and $v_{i}(\bm{r})$ ($i=1,2$) and they satisfy the normalization condition
\begin{align}
\sum_{i=1,2}\int d\bm{r}\left[|u_i(\bm{r})|^2-|v_i(\bm{r})|^2\right]=1.
\end{align}
Substituting $\Psi_{1,2}$ into the time-dependent GP equations lets one obtain the Bogoliubov$-$de Gennes (BdG) equations~\cite{Dalfovo}
\begin{align}
\mathcal{L}[\phi]=\omega\phi,
\end{align}
with $\phi=(u_1,u_2,v_1,v_2)^T$ and
\begin{align}
\mathcal{L}=
\left( 
\begin{matrix}
H_{\mathrm{SOC}} + A - \mu & B \\ B^\ast & - H_{\mathrm{SOC}} - A^\ast - \mu
\end{matrix}
\right),
\end{align}
with
\begin{align}
A=ng\sigma_0+ng\left(
\begin{matrix}
|\varphi_1|^2 & \varphi^\ast_1\varphi_2 \\ \varphi_1\varphi^\ast_2 & |\varphi_2|^2
\end{matrix}
\right),~
B=ng\left(
\begin{matrix}
\varphi^2_1 & \varphi_1\varphi_2 \\ \varphi_1\varphi_2 & \varphi^2_2
\end{matrix}
\right). \notag
\end{align}

\begin{figure*}[]
\includegraphics[width=6.6in]{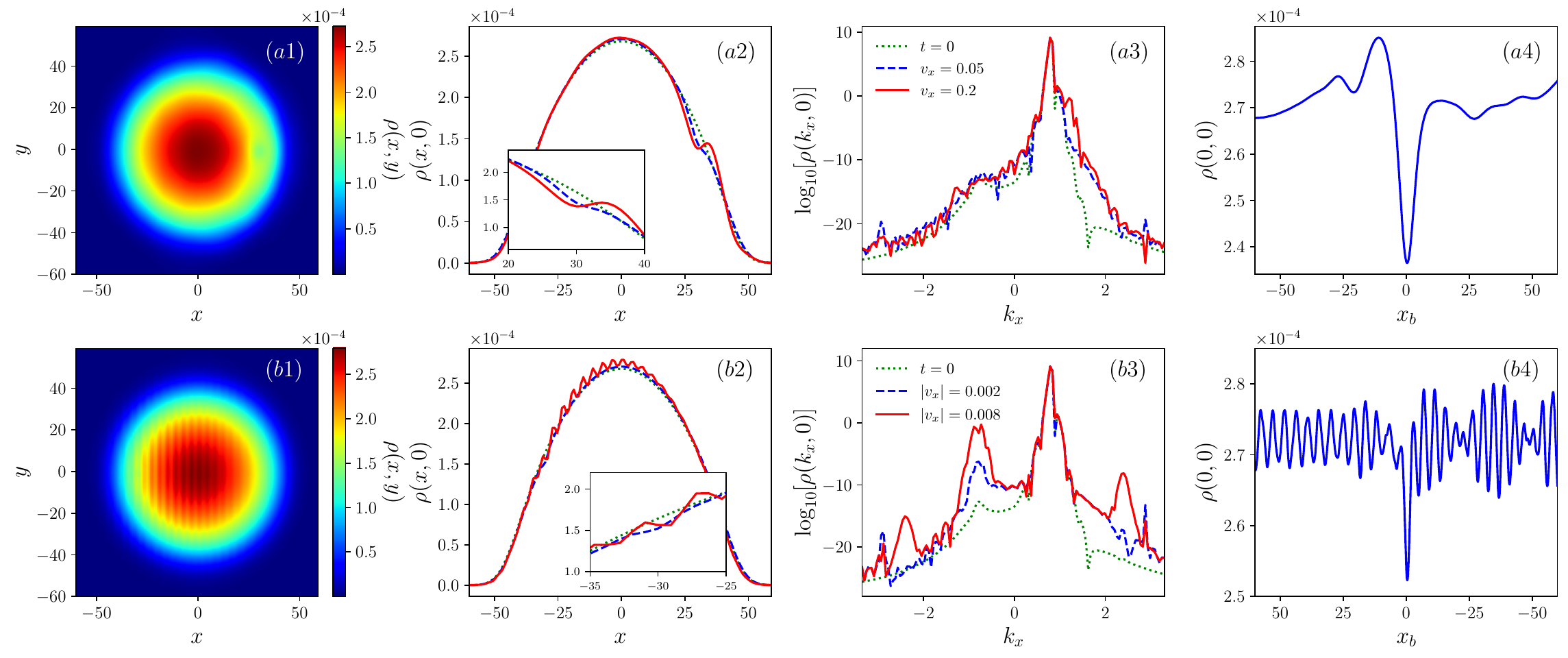}
\caption{Induced dynamics in a spin-orbit coupled BEC  by a weak moving barrier. 
The barrier moves from $x=-60$ to $60$ in (a1)-(a4), and from $x=60$ to $-60$ in (b1)-(b4), while $v_y=0$ in all figures. Snapshots in (a1), (a2) and (a3) are taken at the time where the barrier is at $x=30$; snapshots (b1), (b2) and (b3) are taken when the barrier is at $x=-30$.
(a1) Two-dimensional density of the condensate for $v_x=0.2$. 
(a2)-(a3) Cuts of the condensate and momentum densities  at $y=0$ for different velocities. 
The inset in (a2) shows the densities around $x=30$.
(b1) Two-dimensional density of the condensate for $v_x=-0.008$. 
(b2)-(b3) Cuts of the condensate and momentum densities  at $y=0$ for different velocities. 
The inset in (b2) shows the densities around $x=30$.
(a4) and (b4) show the density at the origin $\rho(0,0)$ as a function of the location of the barrier $x_b$
for $v_x=0.2$ and $v_x=-0.008$,  respectively.
Other parameters are $N=2000$, $\omega_z=0.12$, $\omega_{\perp}=0.006$, and $\Omega=1.2$. }
\label{excitation}
\end{figure*}

The excitation spectrum can be obtained by numerically solving the BdG equations, from which the critical velocities can be calculated. 
Figure~\ref{vc-BdG}(a) shows the excitation spectra along the $x$ direction for different Rabi frequencies.
One can see that the phonon mode softens while the roton mode stiffens with an increasing $\Omega$.
In this parameter regime the system has two degenerate ground states at the quasimomenta  $\pm k$ ($k>0$).
We choose the ground state occupying $-k$ and therefore the roton mode emerges at $q_{\mathrm{rot}}\approx2k$ in the excitation spectrum~\cite{Zhang}.
Note that the roton modes can only be excited along the $+x$ direction, while  phononic excitations can emerge along any directions.
Critical velocities can be determined by the excitation spectrum.
Figure~\ref{vc-BdG} (b) and (c) show the critical velocities of the phonon and roton modes with varying $\Omega$, where the dimensionless density is set as $n=0.025$, corresponding to $1.5\times10^{13}\mathrm{cm}^{-3}$, which is accessible in current experiments~\cite{Yi,Pethick}.

In Fig.~\ref{vc-BdG}(b), we plot the critical velocities (in the unit of $\hbar k_{\mathrm{Ram}}/m$) along the $-x$ direction (green dots) and $y$ direction (red dots) as a function of the Rabi frequency. 
The numerical results show that $|c_{\mathrm{ph},x}|$ decreases with $\Omega$ increasing while $|c_{\mathrm{ph},y}|$ is nearly independent. 
This is due to the one-dimensional character of the Raman-induced spin-orbit coupling.
We also compare the numerical results with the theoretical predictions~\cite{Martone} 
\begin{align}
c_{\mathrm{ph},x}=\sqrt{ng\left(1-\frac{\Omega^2}{4} \right)},~c_{\mathrm{ph},y}\approx\sqrt{ng}.
\label{eq:phonon}
\end{align}
and good agreements can be seen in  Fig.~\ref{vc-BdG}(b). 
The critical velocity of the roton mode as a function of $\Omega$ is shown in Fig.~\ref{vc-BdG}(c). 
Since the roton gap can be estimated as $\Delta\approx ng\Omega^2/4$~\cite{Ji}, one can expect that the slope at the roton minimum is lower than that at the origin, which is clearly confirmed in Fig.~\ref{vc-BdG}(c).
Again, comparing to the analytical expression for the critical velocity along the roton direction ~\cite{Chin}
\begin{align}
c_{\mathrm{rot}}\approx\frac{\Delta}{2k},
\label{eq:roton}
\end{align}
the expected agreement can be found.

These calculations based on the homogeneous system provide reference values of critical velocities for the experimentally relevant systems trapped in an external potential.
However, the inhomogeneous density distribution of a trapped system  is known to have a significant impact on the excitation modes and the critical velocities~\cite{Raman,Stiessberger}. 
In the next section, we will investigate excitations in a spin-orbit coupled BEC trapped in a harmonic trap by dragging a weak Gaussian barrier through it.

\section{Numerical simulations of GP equations}
\label{Sec:III}

Next we consider a harmonically trapped, spin-orbit coupled BEC with a trapping frequency $\omega_z$ in the $z$ direction that is much larger than the ones in the $x-y$ plane, i.e., $\omega_z\gg\omega_{x,y}$. Assuming also that the potential in the $x-y$ plane is symmetric, $\omega_{x,y}=\omega_\perp$, one can describe the dynamics of the system by a two-dimensional, time-dependent GP equation of the form
\begin{align}
i\frac{\partial\Psi}{\partial t}=\left[H^\prime_{\mathrm{SOC}}+Ng^\prime(|\Psi_1|^2+|\Psi_2|^2)+V_{\mathrm{ext}}(\bm{r},t)\right]\Psi,
\end{align}
with $\Psi=(\Psi_1,\Psi_2)^T$ being the two-component wave function. 
The interaction coefficient for the two-dimensional system is given by $g^\prime=2\sqrt{2\pi\omega_z}a_s$ and the external potential is accounted for by $V_{\mathrm{ext}}(\bm{r},t)$ and includes the harmonic oscillator and the moving barrier potentials. 
In the two-dimensional case, the units of the mean density $n_{\mathrm{2D}}$ and interaction strength $g^{\prime}$ are $k^2_{\mathrm{Ram}}$ and $\hbar^2/m$, respectively.
To compare with the homogeneous system, we choose the parameters  $ng=n_{\mathrm{2D}}g^\prime$, where $n_{\mathrm{2D}}$ is determined by the Thomas-Fermi radius~\cite{Pethick}.
Therefore, the excitation spectra along the $x$ direction calculated from the homogeneous BdG equations 
are unchanged for both the three-dimensional and two-dimensional cases. For the calculations we choose $N=2000$, $\omega_z=0.12$ and $\omega_\perp=0.006$, so that we have $n_{\mathrm{2D}}=0.18$ and $g^\prime\approx 155.5$. Finally, the Hamiltonian for the spin-orbit coupling is
\begin{align}
H^\prime_{\mathrm{SOC}}=-\frac{1}{2}\sigma_0\left(\frac{\partial^2}{\partial x^2}+\frac{\partial^2}{\partial y^2}\right)
-i\frac{\partial}{\partial x}\sigma_z+\frac{\Omega}{2}\sigma_x.
\end{align}
To obtain the initial state we consider the barrier to be far away from the condensate and set $V_{\mathrm{ext}}(\bm{r},t)=V_{\mathrm{ho}}(\bm{r})$ with $V_{\mathrm{ho}}(\bm{r})=\frac{1}{2}\omega^2_\perp(x^2+y^2)$ being the harmonic trap potential. 
The ground state of system can then be obtained by using imaginary time evolution. 
	
Next we start to move the barrier through the condensate at a constant velocity for a single pass~\cite{Engels}. 
To avoid complex dynamical phenomena, we consider the barrier to be weak and of a narrow Gaussian shape.
In this situation, the external potential is written as $V_{\mathrm{ext}}(\bm{r},t)=V_{\mathrm{ho}}(\bm{r})+U(\bm{r},t)$, 
where the optical dipole potential is given by
\begin{align}
U(\bm{r},t)=U_0\exp\left[-\frac{(x-x_0-v_xt)^2+(y-y_0-v_yt)^2}{2\sigma^2} \right], \notag
\end{align}
with $U_0$ being the barrier height and $\sigma$ characterizing the width of the barrier. The initial position and the velocity of the barrier are given by $(x_0,y_0)$ and $(v_x,v_y)$,  respectively.
	
To excite the condensate, the barrier velocity should exceed the superfluid critical velocity and we will use the critical velocities calculated for the homogeneous system in Sec.~\ref{Sec:II} as reference values for those of the trapped system which are unknown.

In Fig.~\ref{excitation} we show the results of the real time evolution of the system for different barrier velocities and for fixed Rabi frequency $\Omega=1.2$.
The calculation of the ground state shows that the quasimomentum of the initial state is $k\approx0.8$, which means that the quasimomentum corresponding to the roton excitation is $q_{\mathrm{rot}}\approx-1.6$. 
We move the barrier along the $+x$ and $-x$ directions to generate phonon and roton excitations, respectively.
Figure~\ref{excitation}(a1) shows the condensate density $\rho(x,y)$ for the situation where a barrier of height  $U_0=0.01$ and width $\sigma=2$ moves with $v_x=0.2$ along the anti-roton direction. The snapshot is taken at the time where the barrier is at $x=30$ and one can clearly see a dip in the density distribution at the position of the barrier due to the repulsive nature of the potential. This can be seen more clearly in
Fig.~\ref{excitation}(a2), which shows a cut of the density along $y=0$ for different barrier velocities. 	
At $t=0$ the density $\rho(x,0)$ has a Gaussian-like shape and the associated momentum distribution, shown in Fig.~\ref{excitation}(a3) along the $k_y=0$ direction, has a peak around $k_x=0.8$ (green dotted lines in both plots). Both of these quantities are also shown at the later time where the barrier is located at $x=30$ in Fig.~\ref{excitation}(a2) and (a3), where the blue dashed and red solid lines correspond to the two different velocities $v_x=0.05$ and $v_x=0.2$, respectively.

One can see that for the lower velocity $v_x=0.05$, the cloud stays in the superfluid regime and the distortion to its initial shape is small. 
However, for the larger velocity of $v_x=0.2$ the density is clearly changed and a peak emerges in front of the barrier [see the red solid line in Fig.~\ref{excitation}(a2)], which indicates that a localised wave has been excited by the phonon excitations. 
One can also see that in this situation the peak of $\rho(k_x,0)$ around $k_x=0.8$ is broadened, which is a signal of the excitations. 
In addition, we also plot the time evolution of the density at the origin $\rho(0,0)$ for $v_x=0.2$ in Fig.~\ref{excitation}(a4) as a function of the
location of the barrier, $x_b=v_xt$.
One can see that $\rho(0,0)$ is oscillating with time even after the barrier has left the cloud.
The oscillation period is approximately $R/v_x$, with $R\approx100$ being the spatial extension of the cloud.
We also note that a density dip emerges at $x_b=0$ ($t\approx268$) since the barrier passes the origin and atoms are repelled from the center.

Next we study roton excitations in the system by moving the center of the barrier from $(60,0)$ to $(-60,0)$. 
A snapshot of the condensate density $\rho(x,y)$ for $v_x=-0.008$ at the time when the barrier is at $x=-30$ is shown in Fig.~\ref{excitation}(b1).
One can immediately notice a periodic stripe pattern along the $x$ direction, that is the result of the roton excitation.
The period of the stripes is about 4.05, which agrees with the value of theoretical prediction, $d=\pi/|q_{\mathrm{rot}}|\approx3.92$. 
Figure~\ref{excitation}(b2) and (b3) show the real-space and momentum densities for different $v_x$, again for the time when the barrier is located at $x=-30$.
One can see that the density is only slightly distorted for $v_x=-0.002$, mostly due to the presence of the repulsive barrier [see the inset figure of Fig.~\ref{excitation}(b2)]. 
Furthermore, one can see some additional peaks emerging in the momentum distribution $\rho(k_x,0)$, which, however, 
are very small and do not affect the density profile in  real space. 
Choosing a larger $v_x$, as shown by the red solid lines in Fig.~\ref{excitation}(b2) and (b3), allows one to excite the roton modes and leads to a non-local perturbation of the density despite the local nature of the barrier. 
This is different to the local excitation produced in the phonon excitation case discussed above.
Correspondingly, a peak emerges at $k_x\approx-0.8$ in the momentum density $\rho(k_x,0)$, clearly indicating the excitation of the roton mode. The wave function of the system is therefore in a superposition of plane waves with different quasimomenta, which can be interpreted as a time-varying stripe phase.
This is reminiscent of detection of roton modes in a dipolar gas by using density modulations~\cite{Chomaz},
however here the density-modulated state is an excited state, which does not preserve superfluidity. The density at the origin $\rho(0,0)$ as a function of the location of the barrier is shown in Fig.~\ref{excitation}(b4).
In contrast to the phonon case, $\rho(0,0)$ has two oscillation periods after the barrier leaves the origin.
One period is approximately $d/|v_x|\approx506$, as a result of the spatially periodity of the density, and the other one is about $R/|v_x|\approx12500$, which is associated with boundaries of the cloud. 
	
\begin{figure}[tb]
\includegraphics[width=2.5in]{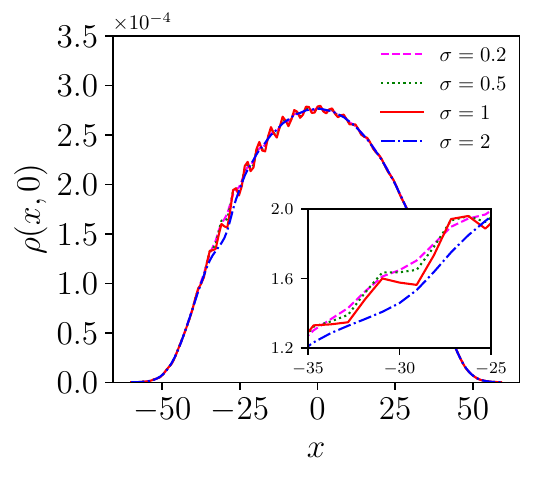}
\caption{Snapshot of the density of the condensate along $y=0$ when the barrier is at $x=-30$ for different barrier widths. 
All other parameters are same as in Fig.~\ref{excitation}(b1).}
\label{density-width}
\end{figure}

The numerical simulations also show that roton modes can only be excited when the barrier width is comparable with the width of a single stripe and in Fig.~\ref{density-width} we show densities for the same parameters as used in Fig.~\ref{excitation}(b1) but for different barrier widths. 
One can see that no excitations appear for barriers of widths $\sigma=0.2,0.5$ and 2. 
For $\sigma=2$, the momentum distribution of the barrier is narrow so that the roton modes can not be excited.
On the other hand, if the barrier width is very small, the excitation is so weak that the contrast of the stripes is reduced.
From the calculations we find that $\sigma=1$ is favourable for the roton excitations.

\begin{figure}[htbp]
\includegraphics[width=2.5in]{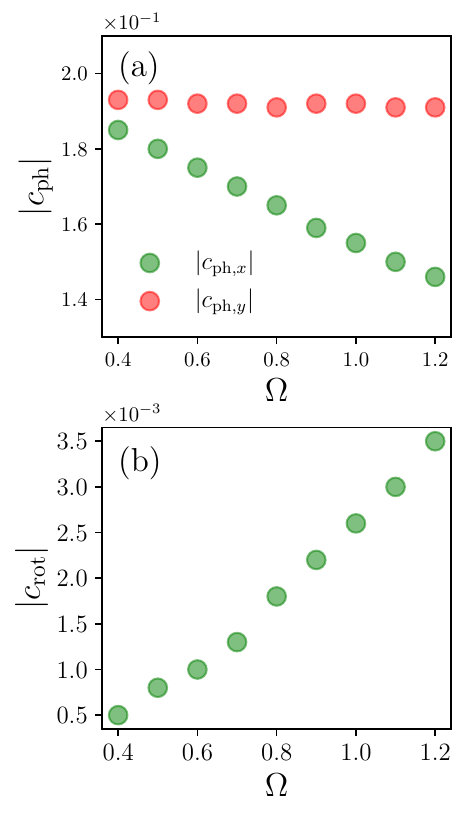}
\caption{Critical velocity of the spin-orbit coupled BEC along directions where no roton exists (a) and along the roton direction (b) obtained by the moving-barrier method. 
The amplitude of the Gaussian barrier is fixed as $U_0=0.01$ while the its width is set as $\sigma=2$ in $(a)$ and $\sigma=1$ in (b). Other parameters are same as in Fig.~\ref{excitation}.}
\label{vc-barrier}
\end{figure}

All the above shows that phonon and roton modes can be excited in a spin-orbit coupled BEC by dragging a barrier through it.
This method can also be used to determine the critical velocities along different directions~\cite{Chin}.
We note that the velocities chosen in Fig.~\ref{excitation} are beyond the critical velocities and they lead to a visible amount of excitations.
In Fig.~\ref{vc-barrier}(a) we show the critical velocities along the anti-roton direction and orthogonal to it as a function of the Rabi frequency.
After preparing the ground states,
we move the barrier along the $+x$ and $+y$ directions to excite phonon modes in the condensate.
For every point in the figure, we choose barriers with different speeds and observe the time-varying density. 
The critical velocities $c_{\mathrm{ph},x}$ and $c_{\mathrm{ph},y}$ are then determined by noting when localised waves emerge in the cloud. 
By increasing $\Omega$, we find that $|c_{\mathrm{ph},x}|$ declines and $|c_{\mathrm{ph},y}|$ is nearly unchanged, which is in accordance with Fig.~\ref{vc-BdG}. 
Here we only show the absolute values of the critical velocities, which can be positive or negative and depend on the quasimomentum of the ground state.
By applying the imaginary time evolution, we find that the system will occupy one of the two degenerate ground states randomly, so that the signs of the quasimomenta can be different if $\Omega$ is changed. 
The critical velocities of the roton modes can be determined by a similar procedure and is shown in Fig~\ref{vc-barrier}(b).
To determine $c_{\mathrm{rot}}$, we choose the emergence of stripes as the signal of roton excitations.
By comparing Fig.~\ref{vc-barrier}(b) and Fig.~\ref{vc-BdG}(b), we find that the magnitudes of $|c_{\mathrm{rot}}|$ for both the two cases agree well, although we obtain larger critical velocities by using the moving-barrier method.

\begin{figure}[htbp]
\includegraphics[width=2.5in]{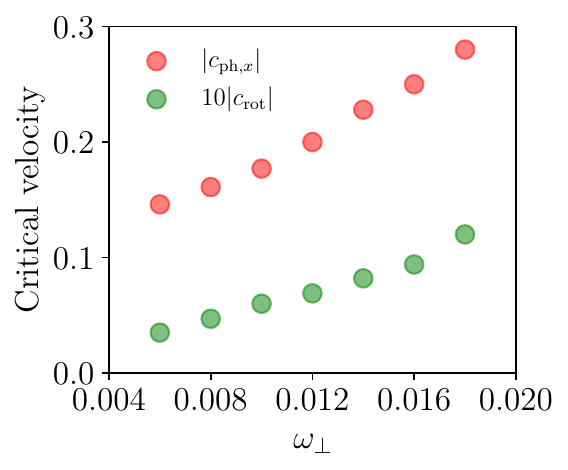}
\caption{Critical velocity of the spin-orbit coupled BEC as a function of the trapping frequency. The Rabi frequency is set as $\Omega=1.2$ and the other parameters are same in Fig.~\ref{excitation}.}
\label{vc-trap}
\end{figure}

While in general the critical velocities calculated by the moving-barrier method are consistent with those obtained from the BdG equations, the former leads to slightly larger values.
This obvious difference is that the BdG equations are based on a homogeneous system while the moving-barrier method is applied to a trapped system. 
This is in contrast to previous works that have pointed out that an inhomogeneous density distribution can lead to a reduced critical velocity~\cite{Raman,Chin}. 
Let us also mention that although the low local density at the edges of the cloud should lead to a reduced critical velocity, the density modulations due to the instability are weak when a low barrier velocity is chosen.

To reduce effect of the inhomogeneity from the start we have in our simulations used narrow Gaussian barriers
with a height that is much less than the chemical potential of the system. This means that a higher speed of the barrier is needed to induce visible excitations.
In our calculations the critical velocity along the non-roton direction is determined by the emergence of a localised wave, while a spatially periodic density is treated as the signal of roton excitations. 
These phenomena are hard to be observed if the barrier speed is close to the value of the homogeneous system and become more obvious when the barrier speed is large.
Therefore critical velocities may be overestimated.

Furthermore, to enhance the accuracy of detecting the critical velocity of the rotons, a larger Rabi frequency can be chosen, since the density period can be raised and this makes the stripes easier to be observed.

The trap parameters can also have an influence on the critical velocity. 
The dependence of $|c_{\mathrm{ph},x}|$ and $|c_{\mathrm{rot}}|$ on the trapping frequency $\omega_{\perp}$ is shown in Fig.~\ref{vc-trap} for a fixed Rabi frequency.
One can see that both $|c_{\mathrm{ph}}|$ and $|c_{\mathrm{rot}}|$ rise with increasing $\omega_\perp$. This can be understood by noting that a
tight confinement of the cloud leads to a enhanced mean density, which requires a higher speed of the barrier to create excitations.
In this situation, the critical velocities of the trapped system move further and further away from the values of a homogeneous system.

\section{Conclusion}
\label{Sec:IV}

In summary, we have studied the generation of phonon and roton excitations in a two-dimensional BEC with Raman-induced spin-orbit coupling by dragging a barrier through it.
The phonon excitations are characterized by the emergence of a localised wave while the roton excitations lead to a periodical modulation of the density distribution.
These phenomena can also be used to determine the critical velocities along different directions that exist as a result of the anisotropic nature of the system.
These studies provide a guide for a practical route to investigtate excitations and dynamics in spin-orbit coupled BECs.

\section*{Acknowledgment}

This work was supported by the Okinawa Institute of Science and Technology Graduate University and used the computing resources of the Scientiﬁc Computing and Data Analysis section. 
Y.Z is supported by National Natural Science Foundation of China with Grants No. 11974235 and No. 11774219.

\bibliography{mbSOC}

\end{document}